\documentclass[prl,twocolumn]{revtex4}
\usepackage{graphicx}
\usepackage{amssymb,amsmath,bm}
\usepackage{enumerate}

\newcommand{\be}{\begin{eqnarray}}
\newcommand{\ee}{\end{eqnarray}}

\begin{document}

\title{MHD turbulence and distributed chaos }

\author{A. Bershadskii}

\affiliation{
ICAR, P.O. Box 31155, Jerusalem 91000, Israel
}

\begin{abstract}

 It is shown, using results of recent direct numerical simulations, that spectral properties of distributed chaos in MHD turbulence with zero mean magnetic field are similar to those of hydrodynamic turbulence. An exception is MHD spontaneous breaking of space translational symmetry, when the stretched exponential spectrum $\exp(-k/k_{\beta})^{\beta}$ has $\beta=4/7$. 
\end{abstract}

\maketitle

\section{Introduction}
  There are two main sources of our interest in the magnetohydrodynamic (MHD) turbulence:  magnetic fusion confinement devices (tokamaks, stellarators etc.) and astrophysics. While the astrophysical observations are full of the scaling spectra (see for a review Ref. \cite{clv}), the measurements in the magnetic fusion confinement devices show the exponential-like spectra \cite{mm},\cite{hom},\cite{s}. The diference is rather deeply rooted. Smooth dynamical systems have exponential like spectra whereas rough ones have scaling spectra. The MHD equations can have both the smooth and the rough attractors (depending on parameters), and even simultaneously but for different sub-ranges of wavenumbers (if the entire range of scales is broad enough, as in the astrophysics). The smooth (chaotic) attractors of the MHD equations attracted much less attention of theoreticians than the rough ones. It could be understood for astrophysics \cite{clv} but not for the practical applications (and for the DNS). 
  
\section{Distributed chaos}  

For an incompressible fluid the MHD equations in Alfv\'enic units have a standard form:
$$
 \frac{\partial {\bf u}}{\partial t} = - {\bf u} \cdot \nabla {\bf u} 
    -\frac{1}{\rho} \nabla {\cal P} - [{\bf B} \times (\nabla \times {\bf B})] + \nu \nabla^2  {\bf u} + {\bf f}_u  \eqno{(1)}
$$
$$
\frac{\partial {\bf B}}{\partial t} = \nabla \times ( {\bf u} \times
    {\bf B}) +\eta \nabla^2 {\bf B} + {\bf f}_m \   \eqno{(2)} 
$$
where the magnetic field ${\bf B}$ normalized by $\sqrt{\mu_0\rho}$ has the same dimensionally as the velocity field ${\bf u}$, and the forcing terms are ${\bf f_u}$ and  ${\bf f_m}$. Magnetic Prandtl number is $P_m =\nu/\eta$.
\begin{figure}
\begin{center}
\includegraphics[width=8cm \vspace{-1cm}]{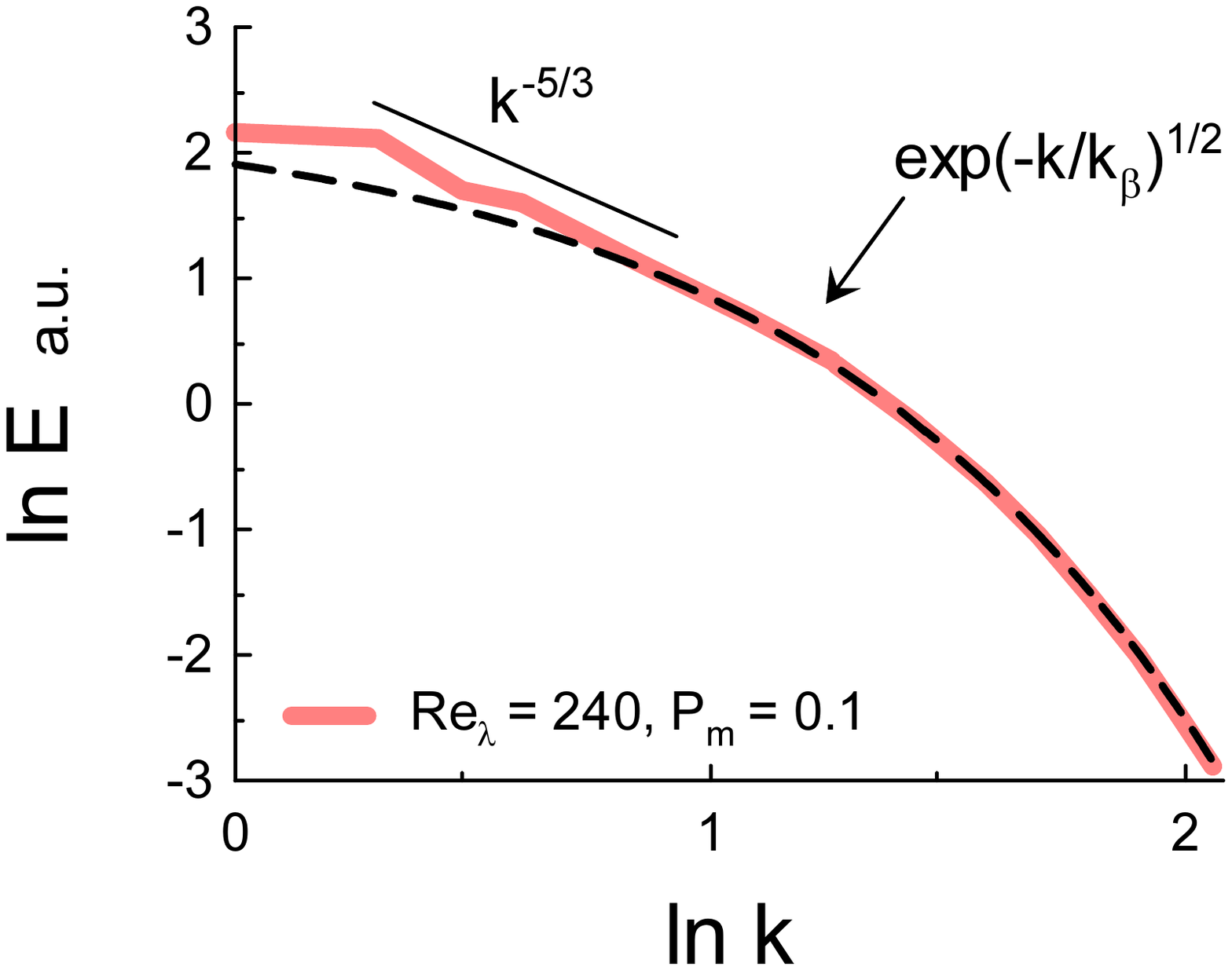}\vspace{-3.7cm}
\caption{\label{fig1} Total energy spectrum of statistically steady (forced) 3D MHD
turbulence. The dashed line is drawn to indicate the stretched exponential spectrum Eq. (3) with  $\beta  =1/2$. }
\end{center}
\end{figure}
  Using results of a direct numerical simulation (DNS) \cite{a} it was noted in Ref. \cite{b1} that for the external stirring force ${\bf f_u}$ taken in the form of a Taylor-Green flow (${\bf f_m}=0$, $P_m=1$, $Re_{\lambda} = 186$, $Re_{\lambda,m}=144$) the distributed chaos range of the wave numbers obeys the same stretched exponential spectral law 
$$
E(k) \propto \exp -(k/k_{\beta})^{\beta} \eqno{(3)}
$$ 
with $\beta =3/4$ as for the isotropic homogeneous turbulence (see also Fig. 2 below).

   For the homogeneous isotropic turbulence velocity correlation integral (Birkhoff-Saffman invariant \cite{saf},\cite{d})
$$
I_2 =\int  \langle {\bf u} \cdot  {\bf u'} \rangle d{\bf r} \eqno{(4)}
$$   
where ${\bf u} = {\bf u} ({\bf x},t)$ and  ${\bf u'} ={\bf u} ({\bf x} + {\bf r},t) $,
dominates the distributed chaos so that the group velocity of the waves driving the chaos
$$
\upsilon (\kappa ) \propto I_2^{1/2}~\kappa^{\alpha} \eqno{(5)}
$$
with $\alpha =3/2$ from the dimensional considerations. The asymptotic theory developed in the Ref. \cite{b1} relates the $\alpha$ to $\beta$ in Eq. (3)
$$
\beta =\frac{2\alpha}{1+2\alpha}   \eqno{(6)}
$$ 
providing the value $\beta =3/4$.
\begin{figure}
\begin{center}
\includegraphics[width=8cm \vspace{-1.9cm}]{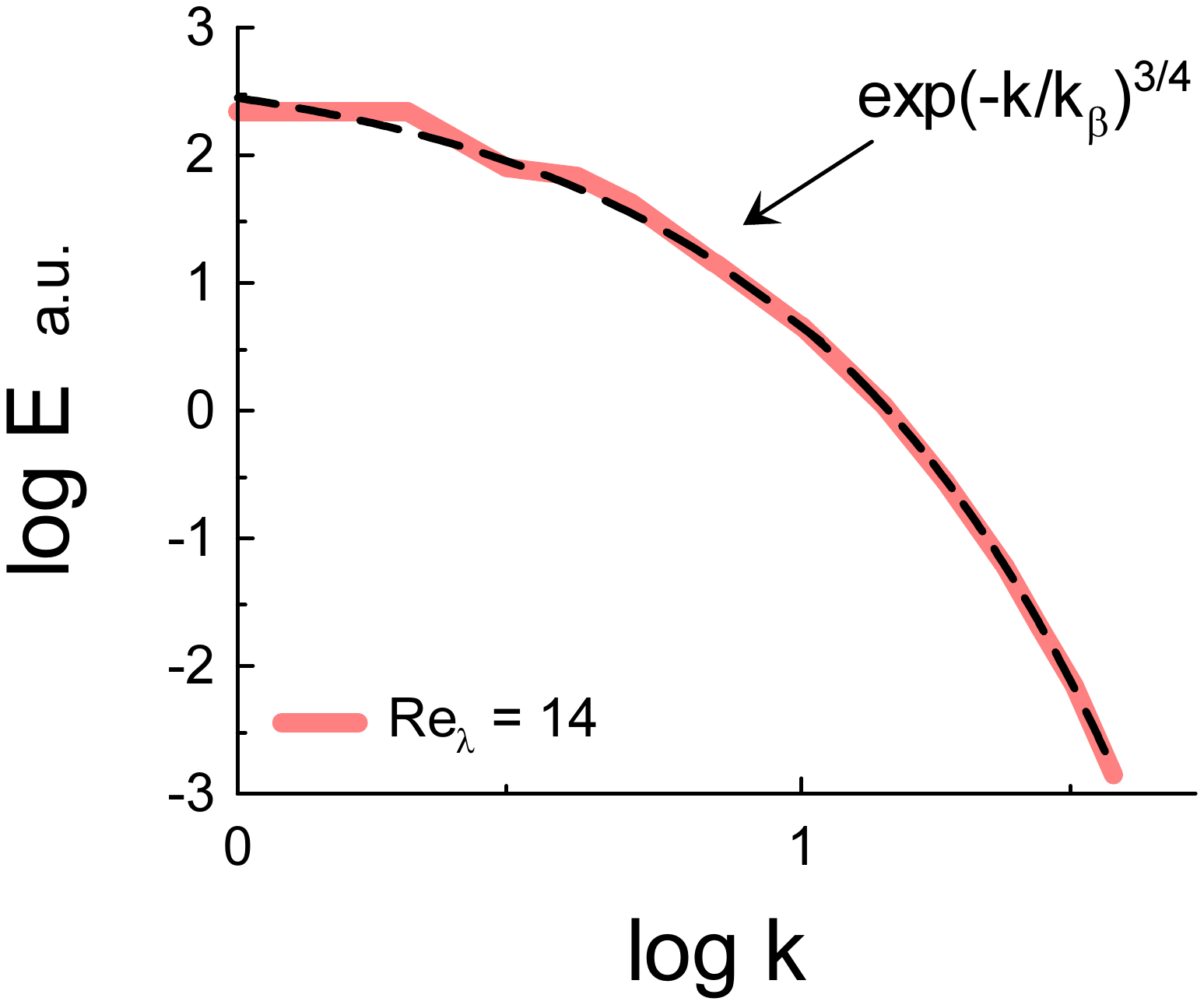}\vspace{-2.4cm}
\caption{\label{fig2}Total energy spectrum of a decaying 3D MHD
turbulence. The dashed line is drawn to indicate the stretched exponential spectrum Eq. (3) with  $\beta  =3/4$.} 
\end{center}
\end{figure}
\begin{figure}
\begin{center}
\includegraphics[width=8cm \vspace{-2cm}]{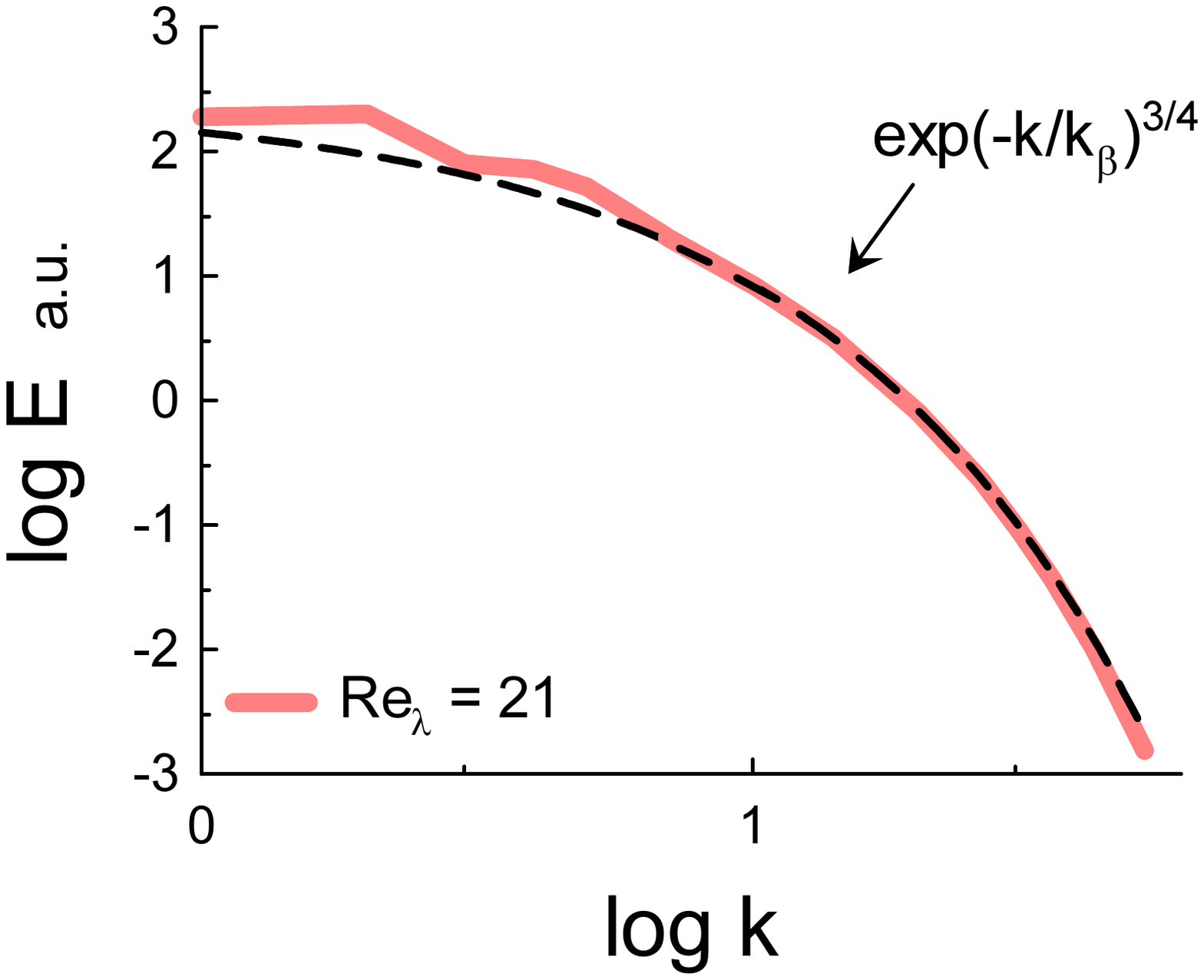}\vspace{-2.5cm}
\caption{\label{fig3} As in Fig. 2 but for $Re_{\lambda} = 21$.}
\end{center}
\end{figure}

\begin{figure}
\begin{center}
\includegraphics[width=8cm \vspace{-1.43cm}]{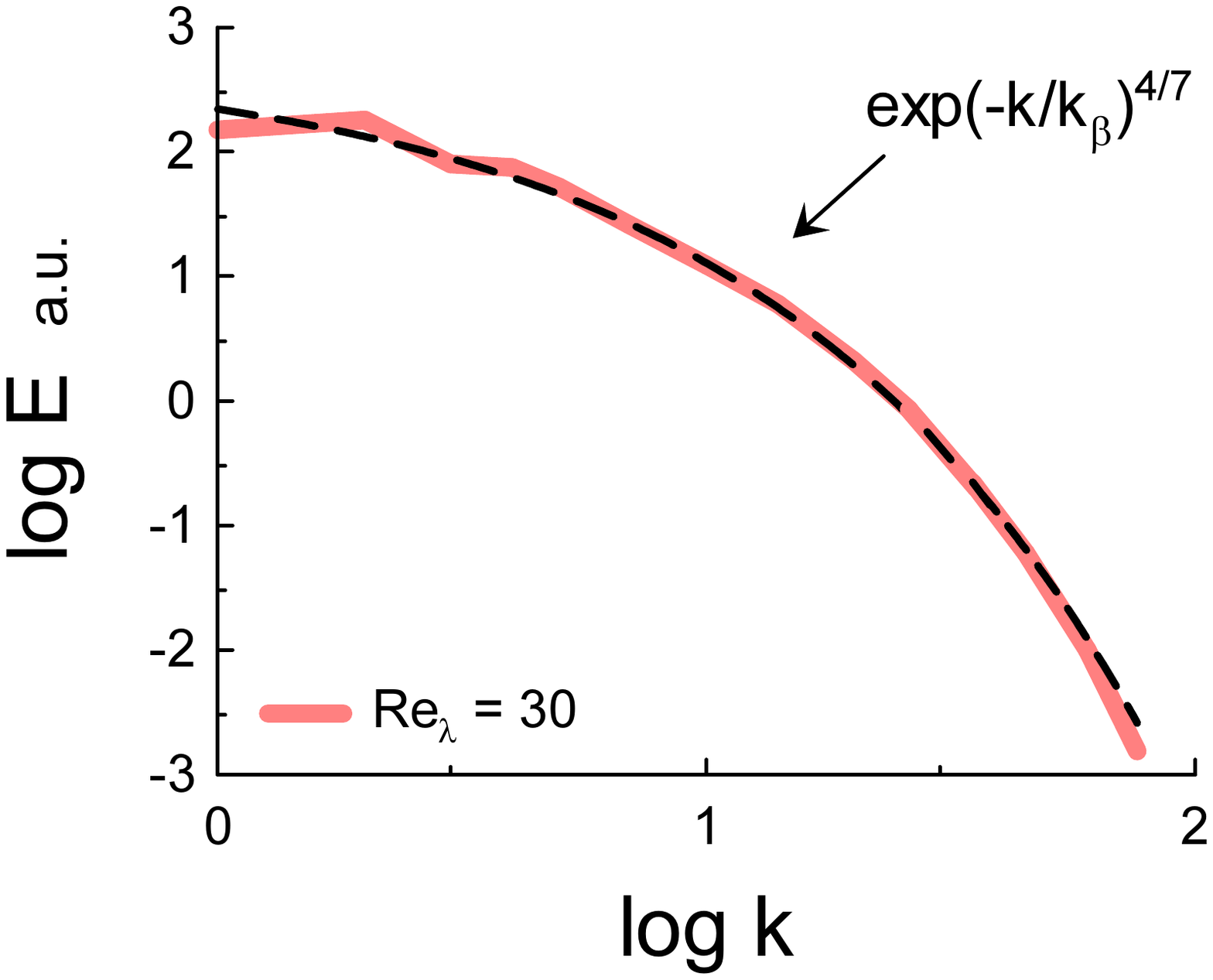}\vspace{-3.35cm}
\caption{\label{fig4} Total energy spectrum of a decaying 3D MHD
turbulence at $R_{\lambda} =30$. The dashed line is drawn to indicate the stretched exponential spectrum Eq. (3) with  $\beta  =4/7$. } 
\end{center}
\end{figure}
\begin{figure}
\begin{center}
\includegraphics[width=8cm \vspace{-1.6cm}]{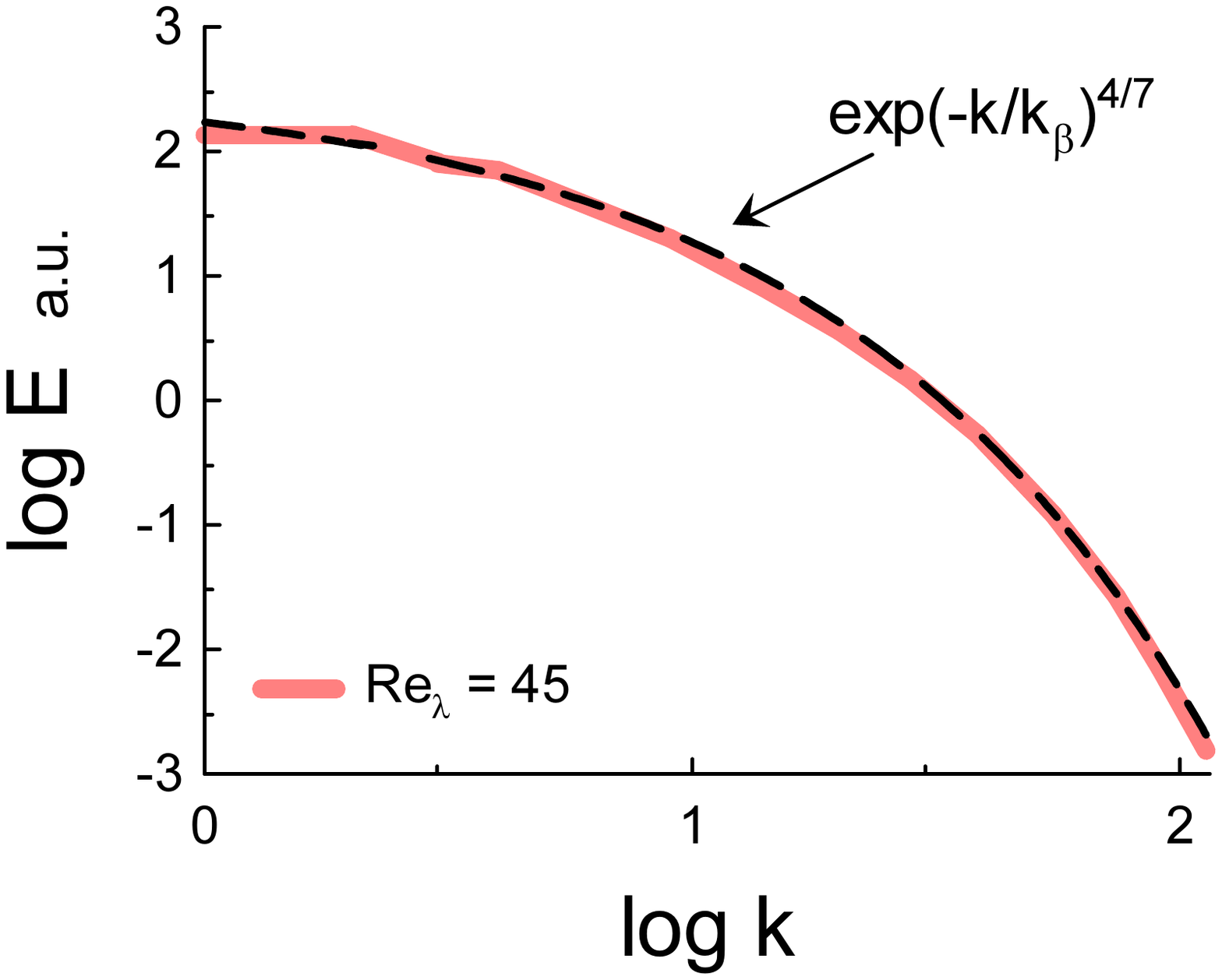}\vspace{-2.8cm}
\caption{\label{fig5}  As in Fig. 4 but for $Re_{\lambda} = 45$.} 
\end{center}
\end{figure}
  According to Noether's theorem \cite{ll} the Birkhoff-Saffman (momentum) invariant is a consequence of space translational symmetry (homogeneity). Therefore, spontaneous breaking of the symmetry switches the domination to a vorticity correlation integral \cite{b2}. 
$$
\gamma = \int_{V} \langle {\boldsymbol \omega} \cdot  {\boldsymbol \omega'}  \rangle_{V}  d{\bf r}  \eqno{(7)}
$$
Substitution the $\gamma$ into Eq. (5) instead of $I_2$ results in $\alpha = 1/2$ and consequently in $\beta =1/2$.
  
    Figure 1 shows total energy spectrum of statistically steady (forced) 3D MHD
turbulence ($P_m = 0.1$, $R_{\lambda}=240$). The DNS data were taken from Fig. 2d of the Ref. \cite{gib}. In this paper
$$
R_{\lambda} = \frac{u_{rms}}{\nu} \left[\frac{<{\bf u}^2+{\bf B}^2>_V}{<\boldsymbol \omega^2 +{\bf j}^2>_V }\right]^{1/2}  \eqno{(8)}
$$
where ${\bf j} = \nabla \times {\bf B}$.  The dashed line is drawn in order to indicate the stretched exponential spectrum Eq. (3) with the above mentioned value $\beta =1/2$ (domination of the vorticity correlation integral Eq. (7)) in the distributed chaos range of wavenumbers $k$. It was noted in the Ref. \cite{b2} that the experimental data indicates such distributed chaos at the plasma edges of some stellarators and tokamaks (see Fig. 9 of the Ref. \cite{b2}).

\section{MHD spontaneous symmetry breaking}

   The spontaneous breaking of the space translational symmetry described in the previous Section was a hydrodynamic one in its nature. At certain conditions an MHD spontaneous breaking of the space translational symmetry (related to the third term in the right hand side of the Eq. (1)) is possible. Let us consider a decaying MHD turbulence (${\bf f_u} ={\bf f_m}=0$) for simplicity. Then, using a consideration analogous to that made in the Ref. \cite{b2} we obtain from the Eq. (1)
$$
\frac{d\int_{V} \langle {\bf u}  \cdot  {\bf u'} \rangle_{V} d{\bf r}}{d t}=-2\nu\gamma -G \eqno{(9)}
$$  
where
$$
G = \int_{V} \langle {\bf u} \cdot  ({\bf B'}\times {\bf j'}) + {\bf u'} \cdot  ({\bf B}\times {\bf j})  \rangle_{V}  d{\bf r}  \eqno{(10)}
$$
Substituting $G$ into the Eq. (5) instead of the $I_2$ we obtain from the dimensional considerations
$$
\upsilon (\kappa ) \propto ~|G|^{1/3}~\kappa^{2/3} \eqno{(11)}
$$ 
i.e. $\alpha = 2/3$, and from Eq. (6) $\beta =4/7$.     

    Figures 2-5 show this transition for a decaying MHD turbulence due to increase of the $R_{\lambda}$ (Eq. (8)) at $P_m=1$. The DNS data were taken from Fig. 2a of the Ref. \cite{gib}. 
Fig. 2 shows total energy spectrum for $R_{\lambda} = 14$. The entire spectrum is covered by the distributed chaos spectrum Eq. (3) with $\beta =3/4$ (situation before the spontaneous symmetry breaking, see previous Section). Fig. 3 shows total energy spectrum for $R_{\lambda} = 21$. One can see a disruption of the isotropic homogeneous dynamics at the large scales (small $k$). Fig. 4 shows total energy spectrum for $R_{\lambda} = 30$. The transition to the $\beta =4/7$ (the MHD spontaneous symmetry breaking) is already realized for the most of the wavenubers except the largest ones. Fig. 5 shows total energy spectrum for $R_{\lambda} = 45$. The MHD spontaneous symmetry breaking is completed. 
\begin{figure}
\begin{center}
\includegraphics[width=8cm \vspace{-1.2cm}]{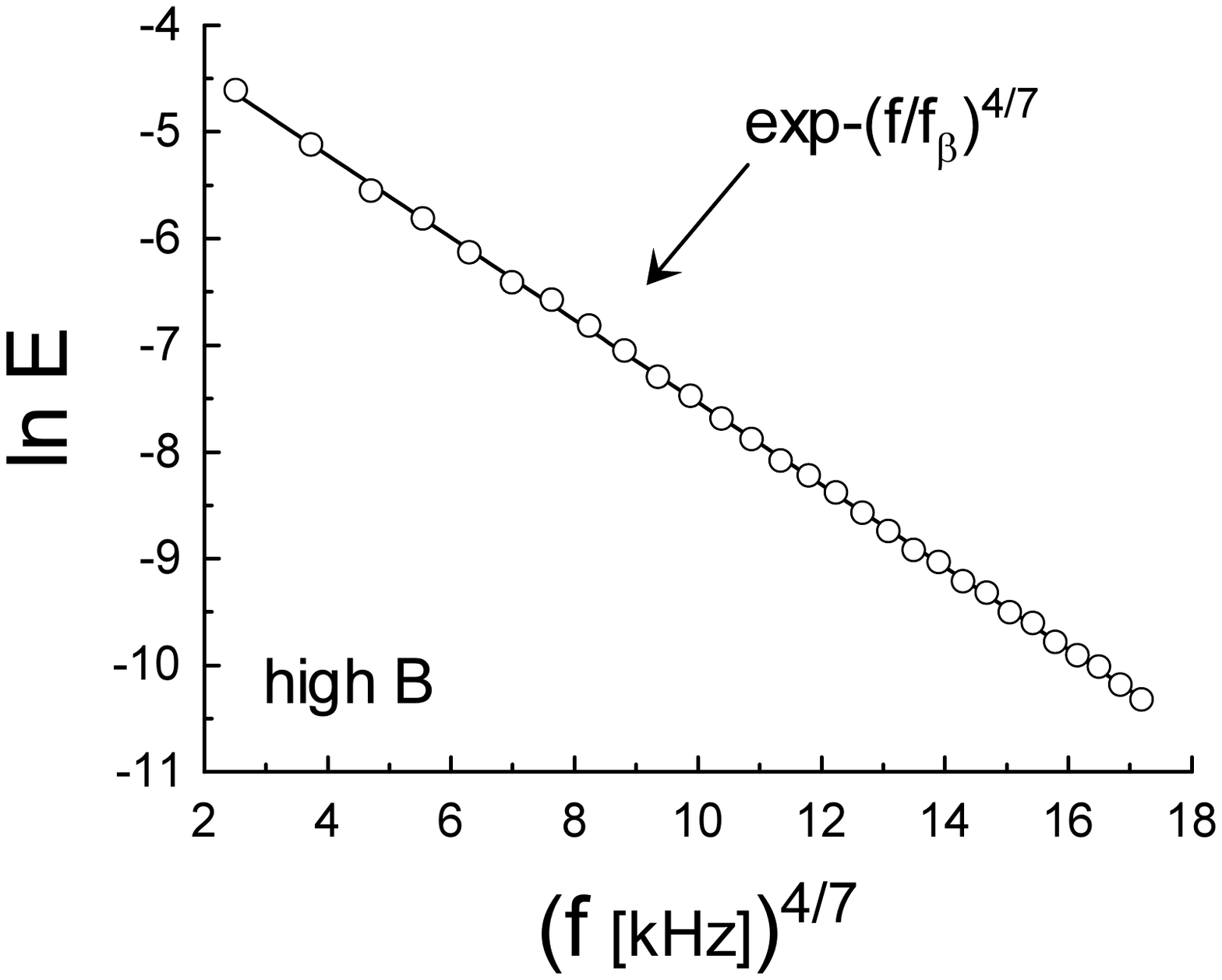}\vspace{-3.35cm}
\caption{\label{fig6}  Power spectrum of ion saturation current in the edge region of TJ-K stellarator at high magnetic field. The scales are chosen to show the stretched exponential Eq. (3) with $\beta =4/7$ as a straight line. } 
\end{center}
\end{figure}
\begin{figure}
\begin{center}
\includegraphics[width=9cm \vspace{-2.1cm}]{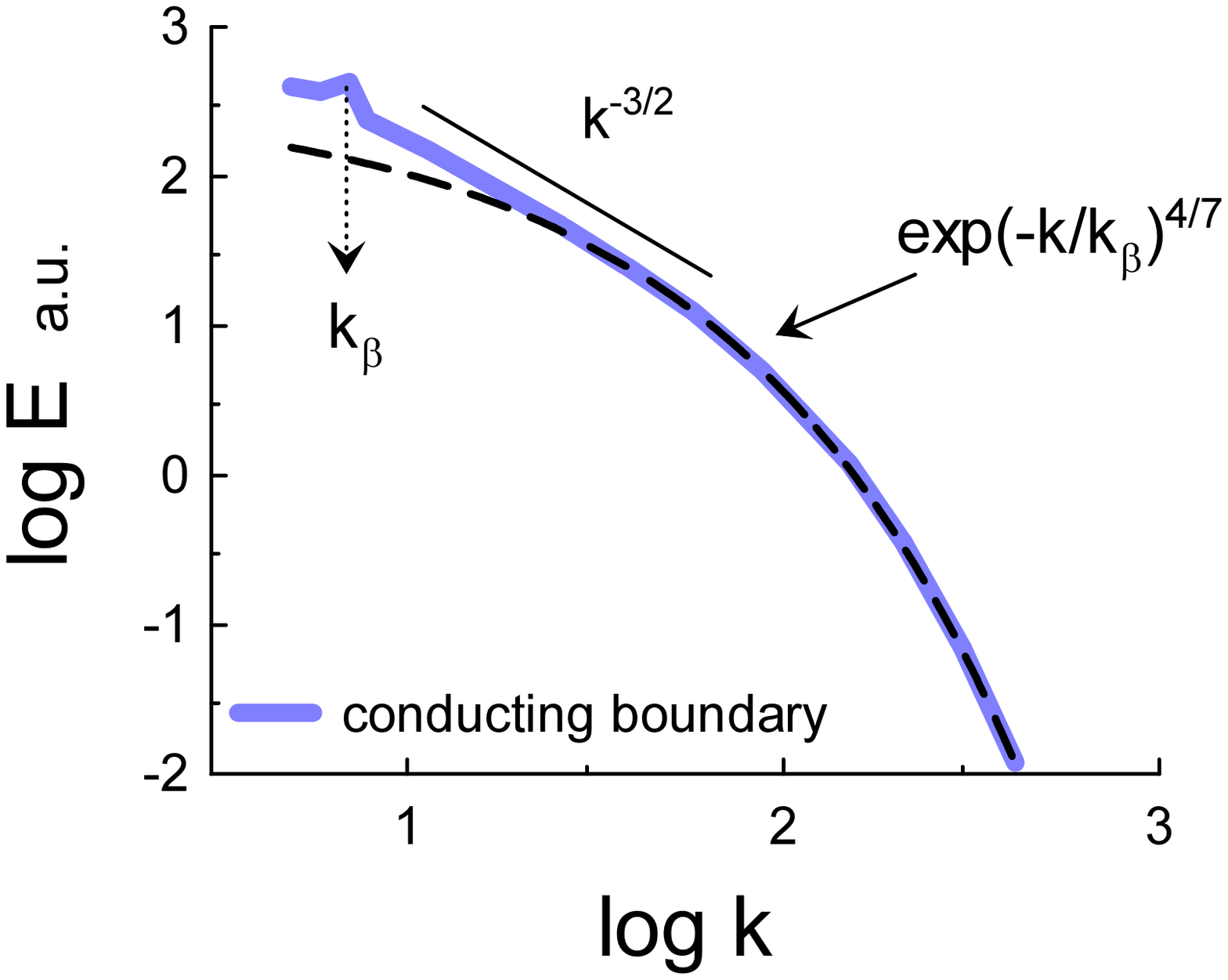}\vspace{-3cm}
\caption{\label{fig7} Total energy spectrum of a Taylor-Green 3D MHD
turbulence (conducting boundary conditions). The dashed line is drawn to indicate the stretched exponential spectrum Eq. (3) with  $\beta  =4/7$.} 
\end{center}
\end{figure}

\begin{figure}
\begin{center}
\includegraphics[width=8cm \vspace{-1.5cm}]{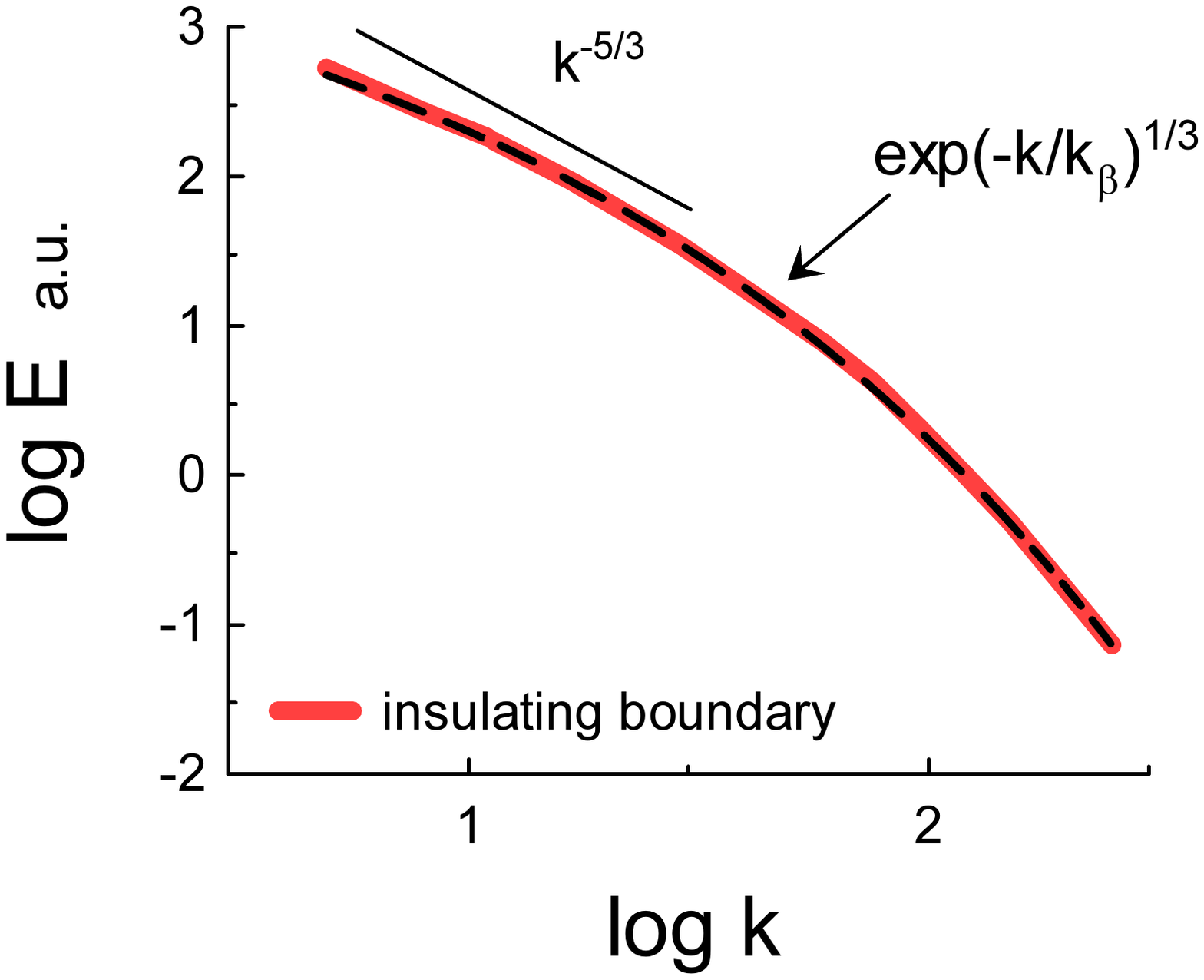}\vspace{-3.6cm}
\caption{\label{fig8} The same as in Fig. 7 but for the insulating boundary conditions. The dashed line is drawn to indicate the stretched exponential spectrum Eq. (3) with  $\beta  =1/3.$} 
\end{center}
\end{figure}  
\begin{figure}
\begin{center}
\includegraphics[width=8cm \vspace{-1.5cm}]{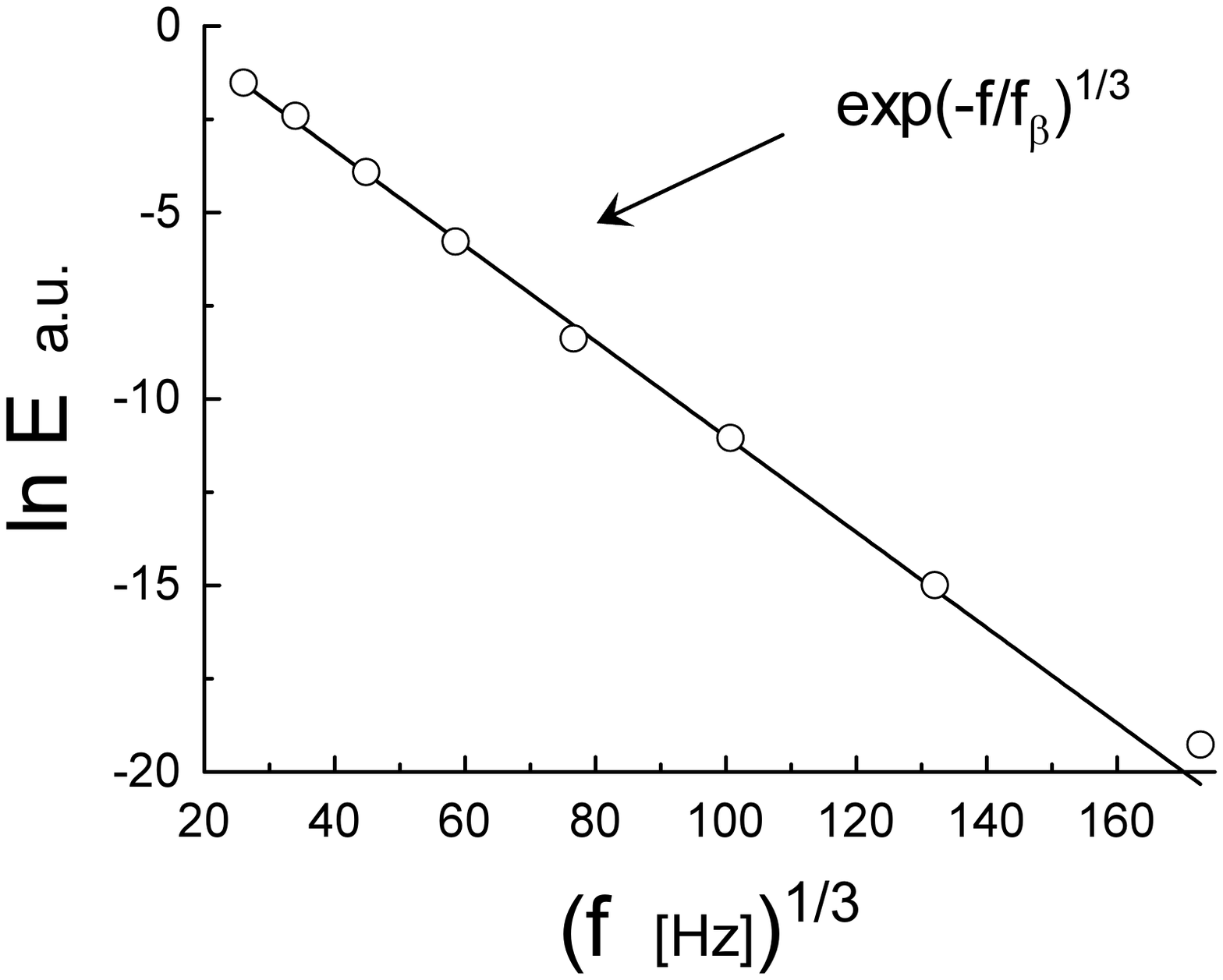}\vspace{-3.6cm}
\caption{\label{fig9}  Total magnetic power spectrum measured in turbulent plasma (a recent Swarthmore Spheromak Plasma Experiment \cite{sbl}). The solid straight line is drawn to indicate the stretched exponential spectrum Eq. (3) with  $\beta  =1/3.$} 
\end{center}
\end{figure}  
    It can be readily shown that in presence of a strong mean magnetic field solitons related dissipation in MHD turbulence with spontaneously broken reflexional symmetry \cite{bkt} can result in the same value of $\beta =4/7$. To support this result we show in Figure 6 power spectrum of ion saturation current in the edge region of a toroidal device with magnetically confined plasma (TJ-K stellarator). The data were taken from Ref. \cite{hom}. The measurements were performed at 
{\it high} magnetic field (244 mT). The scales in this figure are chosen in order to show the stretched exponential Eq. (3) with $\beta =4/7$ as a straight line.  Corresponding spectrum at low magnetic field (72 mT) exhibits the stretched exponential Eq. (3) with $\beta =1/2$ (cf. Fig. 1).

\section{Taylor-Green forcing}

   In a recent DNS \cite{gib},\cite{kbp} of a statistically stable MHD turbulence the forcing terms ${\bf f}_u$ and ${\bf f}_m$ were simulated by the Taylor-Green vortex and its MHD generalization respectively. This forcing allows to simulate conducting and insulating boundary conditions for the currents orientation with respect to the walls of the fundamental box. 
   
     Figure 7 shows total energy spectrum for {\it conducting} boundary conditions ($R_{\lambda} = 110,  P_m =1$). The data were taken from Fig. 7a of the Ref. \cite{kbp}. The dashed line is drawn in order to indicate the stretched exponential spectrum Eq. (3) with  $\beta  =4/7$ (the MHD spontaneous symmetry breaking). The peak at small wavenumbers corresponds to some large scale coherent structures. The arrow under the peak indicates tuning of the distributed chaos (namely - $k_{\beta}$) to the large scale coherent structures.

   Figure 8 shows total energy spectrum for {\it insulating} boundary conditions ($R_{\lambda} = 100,  P_m =1$). The data were taken from the same DNS \cite{kbp}. The dashed line is drawn in order to indicate the stretched exponential spectrum Eq. (3) with  $\beta  =1/3$ for the entire spectrum. This is also a recognizable value of $\beta$. In a recent Ref. \cite{b3} $\beta =1/3$ was obtained for distributed chaos determined by the helicity correlation integral
$$
I = \int_V \langle h({\bf x},t)~h({\bf x}+{\bf r}, t)\rangle_V d{\bf r} \eqno{(12)}
$$
(Levich-Tsinober invariant \cite{lt},\cite{fl}). Indeed, substitution of the $I$ into Eq. (5) instead of $I_2$ 
$$
\upsilon (\kappa ) \propto ~I^{1/4}~\kappa^{\alpha} \eqno{(13)}
$$
provides $\alpha = 1/4$ from the dimensional considerations, and then $\beta =1/3$ from the Eq. (6). 

   It was noted in the Ref \cite{b3} that such distributed chaos determines the spectra (see Fig. 4 in the Ref. \cite{b3}) at the edges of the magnetically confined plasmas at the Large Plasma Device (a cylindrical magnetized plasma column: 60 cm diameter, 17 m long, with the limiters \cite{s}). 
  
   Figure 9 shows total magnetic power spectrum measured in turbulent plasma (a recent Swarthmore
Spheromak Plasma Experiment - a MHD wind tunnel \cite{sbl}). The data were taken from Fig. 15a of the Ref. \cite{sbl}. The scales in this figure are chosen in order to show the stretched exponential Eq. (3) with $\beta =1/3$ as a straight line. 

\section{Discussion}

   Let us generalize the correlation integral Eq. (4)
$$
\tilde{I}_2 =  \int_V  \langle {\bf u} \cdot  {\bf u'} +{\bf B} \cdot  {\bf B'} \rangle_V d{\bf r} \eqno{(14)} 
$$ 
Then from Eqs. (1) and (2) one obtains for the decaying MHD turbulence in the vein of the Ref. \cite{b2}
$$
\frac{d\tilde{I}_2}{d t}=-2\nu\tilde{\gamma} -\tilde{G}  \eqno{(15)}
$$  
where 
$$
\tilde{\gamma} = \int_{V} \langle {\boldsymbol \omega} \cdot  {\boldsymbol \omega'} + P_m^{-1} {\bf j}\cdot {\bf j}' \rangle_{V}  d{\bf r} \eqno{(16)}
$$
and
$$
\tilde{G} = \int_{V} \langle {\bf u} \cdot  ({\bf B'}\times [\nabla \times {\bf B'}]) + {\bf u'} \cdot  ({\bf B}\times [\nabla \times {\bf B}]) -   
$$
$$   
 - {\bf B} \cdot (\nabla \times [ {\bf u'} \times {\bf B'}]) - {\bf B'} \cdot (\nabla \times [ {\bf u} \times
    {\bf B}]) \rangle_{V} d{\bf r}\eqno{(17)}
$$

    If
$
\lim_{R\to\infty}  \tilde{\gamma} = 0 
$
and
$
 \lim_{R\to\infty}  \tilde{G} =0 
$
(where $R$ is radius of the ball $V$ and the decaying MHD turbulence is isotropic and homogeneous at the  limit $ ~R\to\infty$ ), then $ \lim_{R\to\infty} \tilde{I}_2$ is a MHD invariant related to the space translational symmetry (homogeneity) by the Noether's theorem.   

   Substituting the parameters $\tilde{I}_2$, $\tilde{\gamma}$ and $\tilde{G}$ into Eqs. (5) and (11) instead of the parameters $I_2$, $\gamma$ and $G$ one obtains (from the dimensional considerations) the same values of $\beta=3/4,~1/2,~ 4/7$, but now it is a self-consistent MHD consideration.

\end{document}